\documentclass[aps,epsfig,preprint,floatfix]{revtex4}
\usepackage{color}
\usepackage{epsfig,subfigure,float,floatflt,wrapfig,float}
\usepackage{graphicx}
\newcommand{\ts}{\textstyle}
\newcommand{\Pl}{\partial}

\newcommand{\bee}{\begin{equation}}
\newcommand{\ene}{\end{equation}}
\newcommand{\beea}{\begin{eqnarray}}
\newcommand{\enea}{\end{eqnarray}}

\newcommand{\fpar}[2]{\frac{{\ts \Pl \/ #1}}{{\ts \Pl \/ #2}}}
\newcommand{\nder}[3]{\frac{{\ts d^{#1} \/ #2}}{{\ts d \/ #3^{#1}}}}

\begin{document}
\title{Tearing and Surface Preserving Electron Magnetohydrodynamic Modes in A Current Layer}
\author{Gurudatt Gaur}
\email{gurudatt@ipr.res.in}
\author{Predhiman K. Kaw}
\email{kaw@ipr.res.in}
\affiliation{Institute for Plasma Research, Bhat, Gandhinagar - 382 428, India}
\date{\today}
\begin{abstract}
In this paper, we have carried out linear and nonlinear analysis of tearing and surface preserving modes of two 
dimensional (2D) Electron Magnetohydrodynamics (EMHD). A linear analysis shows that the perturbations parallel to equilibrium 
magnetic field $B_0$ (characteristic tangent hyperbolic spatial profile), driven by the current-gradients, 
lead to two different modes. The first mode is the tearing mode having a non-local behavior which 
requires the null-line in the magnetic field profile. Whereas, the second mode is a surface preserving local 
mode which does not require the null-line in the magnetic field. The quantity $B_0 - B_0^{''}$ should change sign for these 
modes to exist. In nonlinear simulations, for tearing case we 
observe formation of magnetic island at the null-line due to the reconnection of magnetic field lines. However, 
for surface preserving mode, a channel like structure is observed instead of the island structure.
\end{abstract}
\maketitle 
\section{Introduction}  
Stability of electron current layers is a long standing topic in theoretical plasma physics. The typical electron 
current layers are found to be formed in many physical situations like, fast z-pinches \cite{kingsep,zp1,zp2}, 
collision less magnetic reconnection \cite{gordeev,reconn1,reconn2,reconn3,reconn4,reconn5,reconn6,reconn7,reconn8}, 
fast ignition phenomena of laser fusion \cite{tabak,hain_mulsar}, plasma opening devices \cite{pos1,pos2,pos3}, inter 
planetary current-carrying plasmas \cite{inter-planetary} etc. These current layers having equilibrium length scale 
smaller than the ion skin depth are prone to various current-gradient driven instabilities under which they evolve, 
sometimes to the point of complete destruction. In typical electron current layers, current flows faster than the 
Alfven velocity and the Magnetohydrodynamic (MHD) model is not applicable. The 
stability of these current layers has been studied using Electron Magnetohydrodynamic (EMHD) model of plasmas \cite{kingsep}. 
EMHD is a single fluid description of plasma which describes the dynamics of electron only by ignoring the ion response. 
EMHD model has proven to be very convenient in describing numerous phenomena occurring at fast time and short length scale 
where MHD is not applicable. A rich literature on this model is available, which can be found elsewhere. Here, we will 
discuss applicability of this model to describe the current gradient driven instabilities.

\begin{figure}
\begin{center}
\includegraphics[width=24.cm,height=12.cm]{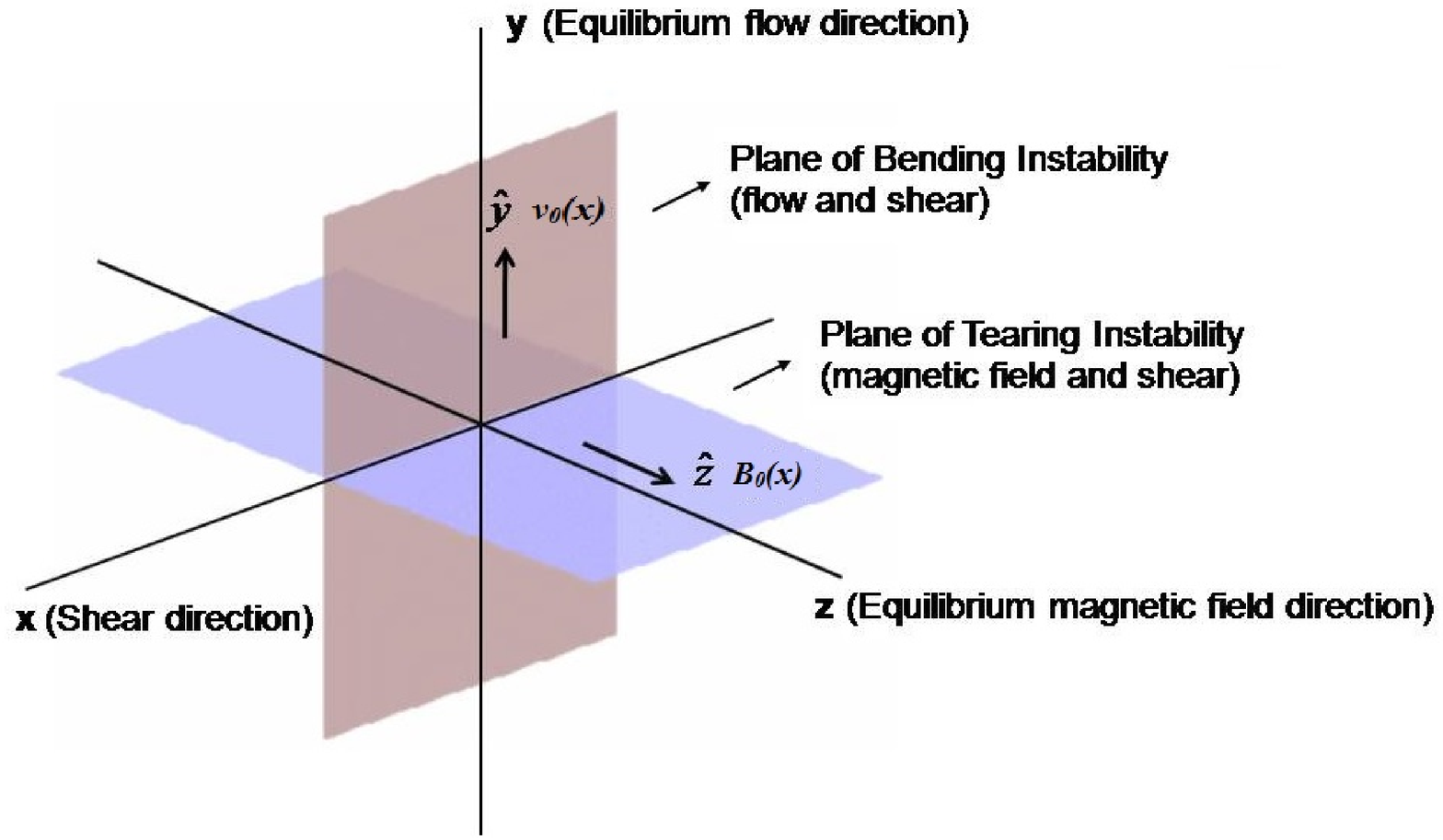}
\end{center}
\caption{}
\label{Fig0}
\end{figure}

The current-gradient driven instabilities in the framework of EMHD have been previously considered by 
Califano et al. \cite{tear_bend}, where they have been broadly categorized as tearing and bending instabilities depending 
upon the orientation of perturbations relative to the equilibrium magnetic field. The classification 
can be understood from Fig.~{\ref{Fig0}}. The perturbations propagating along the direction of flow (perpendicular to 
magnetic field) give rise to the excitation of Kelvin Helmholtz (KH) like modes \cite{das_kaw_kh,gaur1}, which bends the 
flow lines and leads to the formation of vortex structures. This mode is known to play the role in stability of vortices 
generated by the interaction of ultra intense laser pulse with a plasma \cite{pukhov}, generation of small scale 
turbulence \cite{drake_99}, anomalous stopping of the energetic electron beam in fast ignition \cite{jain_pla} etc. 
The other choice of perturbations (i.e. propagating along the magnetic field) gives excitation to the collisionless 
tearing instability \cite{gordeev,reconn1} of thin current sheets which leads to the magnetic reconnection in 
the presence of electron inertia.

Apart from these tearing-bending instabilities, an inertial scale instability is known which also falls in the 
category of current-gradient driven instabilities \cite{jain_kink_lin,lukin,gaur2}. This mode shares the geometry of tearing 
mode [Fig.~\ref{Fig0}], but unlike the tearing mode it is a local mode and does not require reversed equilibrium magnetic field 
configuration. This mode preserves the surface of magnetic flux. Henceforth, this mode shall be referred to as non-tearing mode. 
In the study presented here, we suppress the KH mode by not 
considering the perturbations along the flow direction. Hence our study is two dimensional with perturbations confined in the 
plane containing the magnetic field and gradient directions.

In the literature, the tearing mode has been studied for a 1D magnetic field profile, $B_{0} = tanh(x/\epsilon)$, 
which is a Harris current sheet \cite{harris} with thickness $\epsilon$. 
For this choice of profile, as we will see, one of the conditions for non-tearing mode, $B_0 B_0^{''} > 0$ 
(see \cite{lukin}) is not satisfied and hence the non-tearing mode was not present in the earlier studies of tearing mode. 
In order to study the non-tearing mode Lukin \cite{lukin} used the Harris 
current sheet equilibrium but up-shifted so that $B_0 > 0$ everywhere and the condition $B_0 B_0^{''} > 0$ is satisfied. 
In the studies by Jain et al. \cite{jain_kink_lin} and Gaur 
et al. \cite{gaur2} also, the magnetic profile had the definite sign ($B_0 < 0$ everywhere) and the non-tearing mode was 
present. However, in these studies, due to the absence of null-line in the magnetic field profile, the tearing mode was not 
present. Thus, so far, the two modes have not been investigate simultaneously in the same system. Here, we study the 
two modes simultaneously present in the same system for a suitably tailored 1D magnetic field profile.

The paper has been organized as follows. In section II we briefly discuss the model used and the equilibrium configuration of 
the system. Section III contains linear instability analysis where we discuss the stability conditions of the two modes. 
Section IV presents the results of nonlinear simulations that we carried out to understand the nonlinear state when two 
modes are operative separately and simultaneously. Section V summarizes our work.

\section{Model and Governing Equations}

The EMHD model works for the phenomena involving the fast time scales 
$\omega_{ci,pi}<<\omega<<min(\omega_{pe},\omega_{pe}^2/\omega_{ce})$ and short 
spatial scales $\rho_e,\lambda_D<<\lambda<<\rho_i,d_i$. Here, $\lambda_D$ is the Debye radius, $\omega_c$ and $\omega_p$ are 
the gyro and plasma frequencies respectively and $\rho$ and $d$ denote the larmor radius and skin depth respectively. 
Subscripts $e$ and $i$ represent the electron and ion species of plasma. At these scales the ions can be assumed to be static 
and unmagnetized. Thus, the EMHD equations are obtained from combined set of electron fluid equations and the Maxwell's 
equations. The EMHD model can be described by the following set of dimensionless equations, 
\begin{eqnarray}
\fpar{}{t}(\nabla^2\vec{B}-\vec{B})  &=& \vec{\nabla} \times [\vec{v} \times (\nabla^2\vec{B} - \vec{B})]   \nonumber  \\
\vec{v_e} &=& -\vec{\nabla} \times \vec{B}
\label{emhd_eq}
\end{eqnarray} 
Here, the  length scale has been normalized by electron skin depth $d_{e} = c / \omega_{pe}$, magnetic field by 
a typical magnitude concerning any problem, e.g.  $B_{N}$, the  
 time has been normalized by  the  electron cyclotron period corresponding to the normalizing magnetic field  $B_{N}$.
The first equation represents the evolution of generalized vorticity 
$\vec \nabla \times \{\vec v_e-\vec A\} = \nabla^2\vec B - \vec B$. Here, $\vec{A}$ is the vector potential. Second 
equation is Ampere's law in which displacement current has been ignored by taking, 
$\omega<<min(\omega_{pe},\omega_{pe}^2/\omega_{ce})$. Under this assumption the density fluctuations are ignored 
$(n \sim 0)$ and the condition of quasi-neutrality demands the incompressibility of electron fluid i.e. 
$(\nabla.\vec{v_e} = 0)$. Moreover, the electron ion collisions have also been ignored.

In two dimensions (with variation along $x-z$ only) with the use of $\nabla.\vec{B} = 0$ condition the total magnetic 
field can be expressed as, $\vec{B} = b\hat{y}+\hat{y}\times\nabla\psi$. The corresponding electron velocity would be 
expressed as, $\vec{v_e} = -\nabla\times\vec{B} = \hat{y}\times\nabla b-\hat{y}\nabla^2\psi$. Thus, the above set of 
EMHD equations [Eqs.~ (\ref{emhd_eq})] can be cast in terms of the evolution of two scalars,
\begin{eqnarray}
\fpar{}{t}(\nabla^2  b - b)  + \hat{y} \times \nabla b  \cdot \nabla (\nabla^2  b-b)
 -  \hat{y} \times \nabla \psi  \cdot \nabla (\nabla^2  \psi-\psi) &=& 0 \nonumber \\
\fpar{}{t}(\nabla^2  \psi - \psi)  + \hat{y} \times \nabla b  \cdot \nabla (\nabla^2  \psi - \psi) 
 &=&  0  
\label{2d_emhd}
\end{eqnarray}

The equilibrium of Eq.~(\ref{2d_emhd}) is defined as follows. We choose, $b_0 = constant$. With this and choice of one 
dimensional equilibrium, Eq.~(\ref{2d_emhd}) becomes, 
\begin{eqnarray}
\nder{2}{\psi_0}{x} - \psi_0 &=& F(\psi_0) \\
\Rightarrow\quad\quad\quad\quad \nder{2}{\psi_0}{x} &=& G(\psi_0)
\label{eqb}
\end{eqnarray}
Here, $G(\psi_0) \equiv F(\psi_0) + \psi_0$. $F$ and $G$ are the arbitrary functions. Simplifying Eq.~(\ref{eqb}) further gives, 
\begin{eqnarray}
H(\psi_0) &=& x \\
\Rightarrow\quad\quad\quad\quad \psi_0 &=& f(x) \\
\Rightarrow\quad\quad\quad\quad B_0 &=& - \frac{d \psi_0}{dx} = f'(x)
\end{eqnarray}
Here, $H(\psi_0) = (1/\sqrt{2})\int{I(\psi_0) d\psi_0} - K_2$; $I(\psi_0) = 1/[\int{G(\psi_0 d\psi_0 + K_1)}]^{1/2}$. 
$K_1$ and $K_2$ are the constants and $H$ and $f$ are some arbitrary functions. Thus, we choose an equilibrium sheared magnetic 
field $\vec{B}_0(x) =  B_0(\textit{x})\hat{z}$. 
This corresponds to a sheared electron flow directed along $\hat{y}$ axis as, $\vec{v}_0(x)   =  v_0(\textit{x})\hat{y}$. 
This choice would exclude the KH modes in our system as there are no variations along the equilibrium velocity.

We linearize the EMHD equations [Eqs.~(\ref{2d_emhd})] about the above equilibrium. Since the equilibrium is the function of $x$ 
only, we take Fourier transform in $z$ and $t$ to obtain the following set of linearized equations, 
\begin{eqnarray}
\nder{2}{b_1}{x}-(1+k_z^2)b_1 + 
\frac{k_z B_0}{\omega} \left( \nder{2}{\psi_1}{x} - k_z^2 \psi_1 \right) - \frac{k_z B_0^{\prime\prime}}{\omega} \psi_1&=& 0 \nonumber  \\
\nder{2}{\psi_1}{x} - (1+k_z^2) \psi_1 - \frac{k_z(B_0-B_0^{\prime\prime})}{\omega} b_1 &=& 0
\label{lin_emhd}
\end{eqnarray}

\section{Linear Instability }
In this section we analyze the set of coupled linearized equations (\ref{lin_emhd}) obtained in the previous section to 
understand the growth rate and eigen functions of tearing and non-tearing modes.

We solve the Eqs.~(\ref{lin_emhd}) numerically as a matrix eigen value problem for 
\begin{eqnarray}
 B_0(x) = tanh(x/\epsilon) + C_0
\end{eqnarray}

\begin{figure}
\begin{center}
\includegraphics[width=18.cm,height=12.cm]{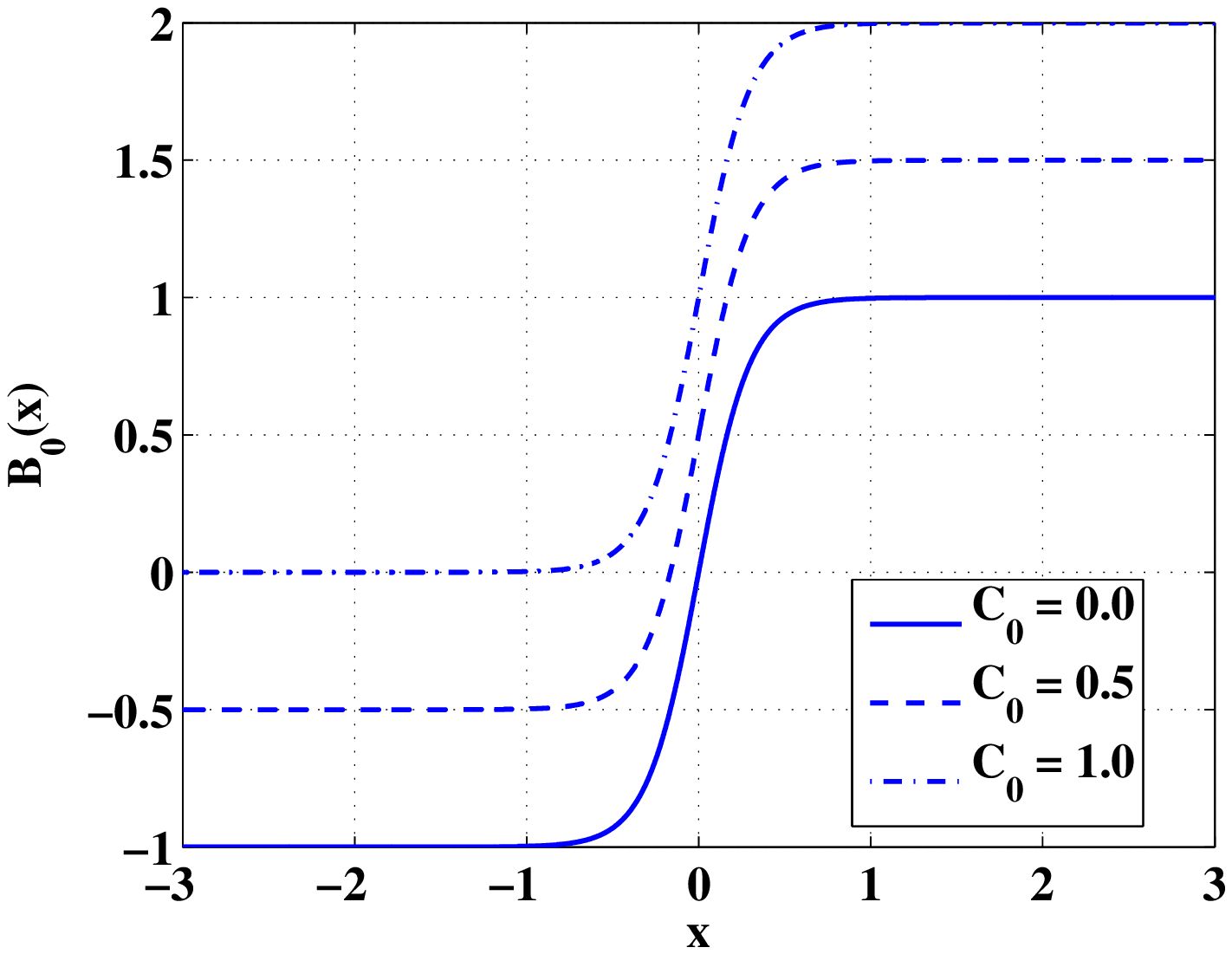}
\end{center}
\caption{}
\label{Fig1}
\end{figure}

Where, $\epsilon$ is shear width and $C_0$ is a uniform magnetic field added externally in the direction of tangent hyperbolic field. 
The presence of this magnetic field does not disturb the equilibrium, however, it shifts/removes the null-point in the profile depending 
upon its magnitude [Fig.~\ref{Fig1}]. We numerically obtain the growth rate as eigen values for different choices of $C_0$. 
We would like to point out here that the modes with value of $C_0 = 0.0$ would be termed as pure tearing modes and the modes with value 
of $C_0 \geq 1.0$ as pure non-tearing modes. Modes with value of $0 < C_0 < 1.0$ would be the mixed modes. We would make this nomenclature clear 
later in this section.

\begin{figure}
\begin{center}
\includegraphics[width=18.cm,height=12.cm]{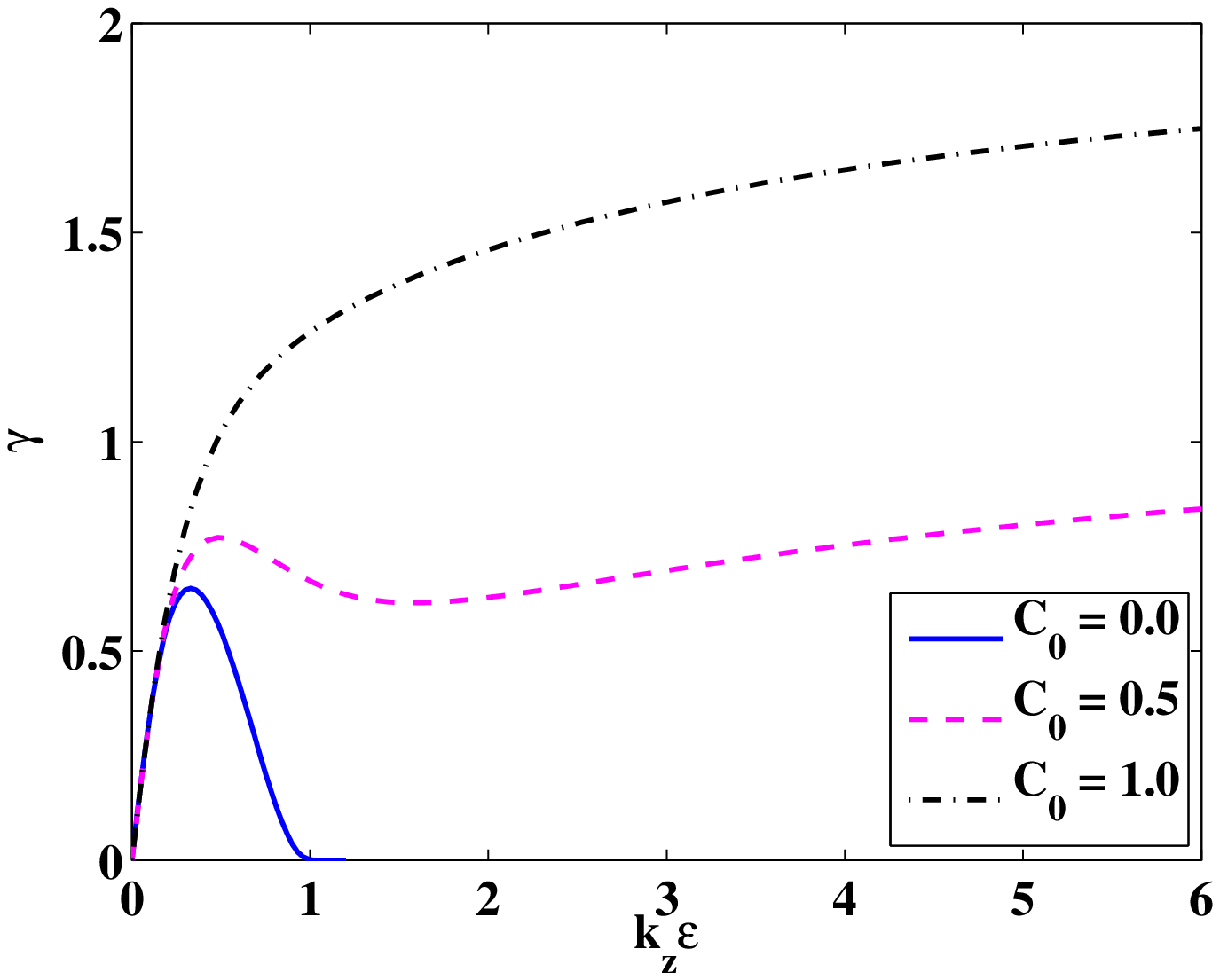}
\end{center}
\caption{}
\label{Fig2}
\end{figure}

In Fig.~\ref{Fig2} we plot the growth rate as a function of $k_z$. The three different curves correspond to different values of $C_0$. 
The growth rate for pure tearing case (i.e. $C_0 = 0$) is nonlocal i.e. only modes with $k_z\epsilon \sim \mathcal{O}(1)$ are unstable. 
All the local modes (with $k_z > 1/\epsilon$) are stable. 
However, as the value of $C_0$ is made finite, local modes are also become unstable. 
For the case $C_0 = 0.5$, the growth rate curve shows a dip and then saturates at higher $k_z$ values.
For the case $C_0 = 1.0$, the growth rate curve shows no dip. This is the case when magnetic field has no null-line and the 
tearing mode is absent. The growth rate curve saturates in the local region and becomes 
independent of $k_z$. This behavior is  consistent with studies of Lukin \cite{lukin}.

\begin{figure}
\begin{center}
\includegraphics[width=18.cm,height=12.cm]{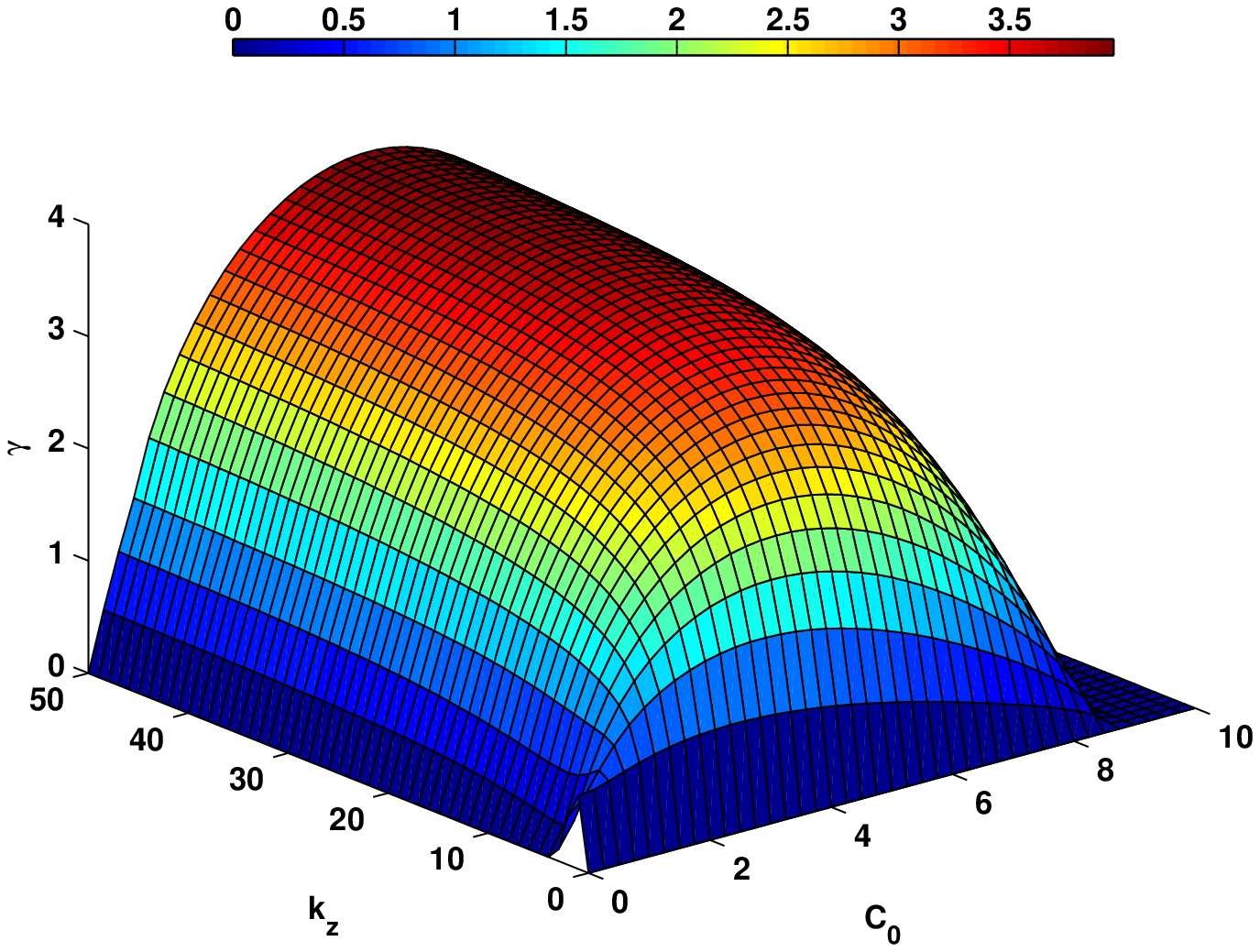}
\end{center}
\caption{}
\label{FigC}
\end{figure}

In Fig.~\ref{FigC}, we show the surface plot of growth rate as a function of $k_z\epsilon$ and $C_0$. 
This plot shows that the growth rate of local modes first increases with $C_0$ and then vanishes for some $C_0 = C_0^{stable}$.

\begin{figure}
\begin{center}
\includegraphics[width=18.cm,height=12.cm]{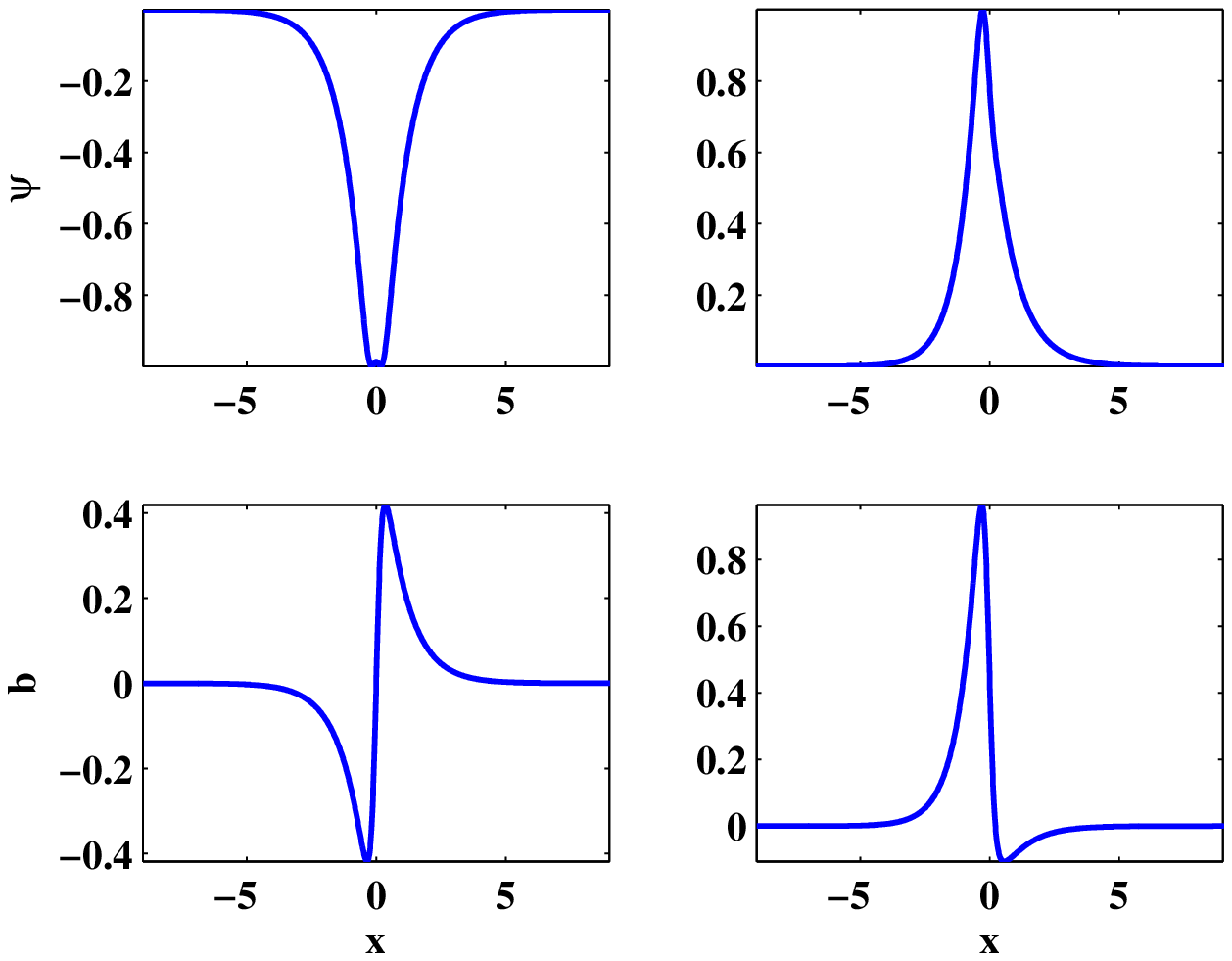}
\end{center}
\caption{}
\label{Eigfn}
\end{figure}

In Fig.~\ref{Eigfn} we plot the eigenfunctions of pure tearing and pure non-tearing modes in the left and right panels 
respectively. The eigenfunction of tearing mode shows its standard spatial character, where $\psi$ is even in $x$ and slowly varying 
around $x = 0$, while $b$ is odd in $x$ and peaked around $x = 0$. Whereas, the eigen mode structure loses the symmetry for pure 
non-tearing mode $C_0$. The structure is asymmetric around $x = 0$ both in $\psi$ and $b$.

\subsection*{Investigation of local region}
\begin{figure}
\begin{center}
\includegraphics[width=18.cm,height=12.cm]{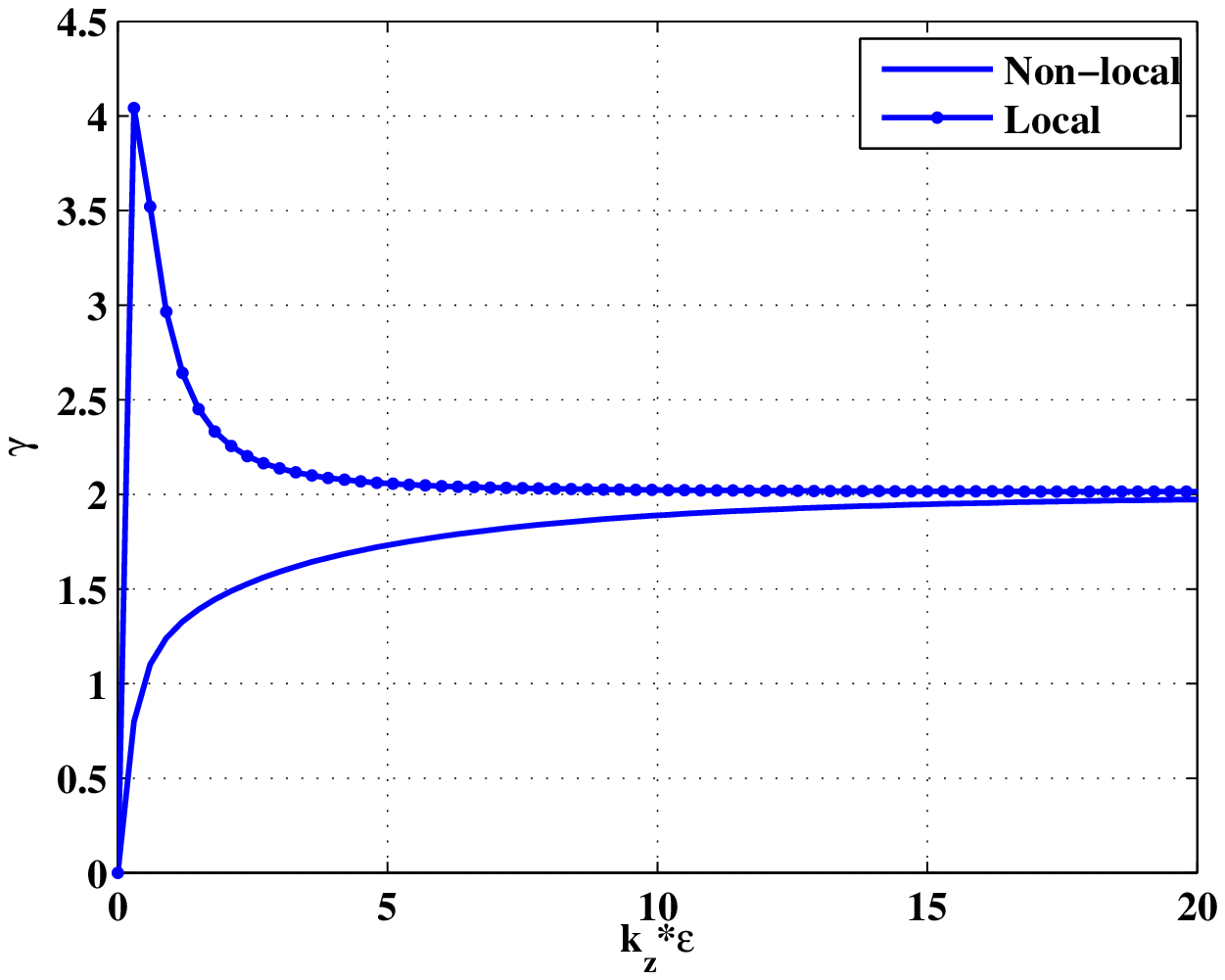}
\end{center}
\caption{}
\label{Fig3}
\end{figure}

Assuming that the perturbation scales ($k_z^{-1}$) are sharper than the equilibrium scales, we can take the Fourier 
transform of Eqs.(\ref{lin_emhd}) along  $x$ also and obtain the dispersion relation as follows, 
\begin{eqnarray}
\omega^2 (1+k_0^2) = k_z^2(B_0-B_0^{\prime\prime})(k_0^2 B_0+B_0^{\prime\prime})
\label{loc_dp}
\end{eqnarray}
Here $k_0 = (k_x^2+k_z^2)^{1/2}$. For instability, RHS should be negative. From the dispersion relation it is clear 
that for $B_0^{\prime\prime} (=-v_0^\prime) = 0$, there is no instability, which implies that the modes are current-gradient 
driven (in EMHD the current is directly proportional to electron velocity through the relation $\vec{J} = -ne\vec{v}$). 
Using the above dispersion relation [Eq.~(\ref{loc_dp})], we obtain a growth rate curve which matched the non-local growth rate 
curve at high $k_z$ values [Fig~\ref{Fig3}]. The mismatch at small $k_z$ value this is because the local analysis is not valid there.

\begin{figure}
\begin{center}
\includegraphics[width=18.cm,height=12.cm]{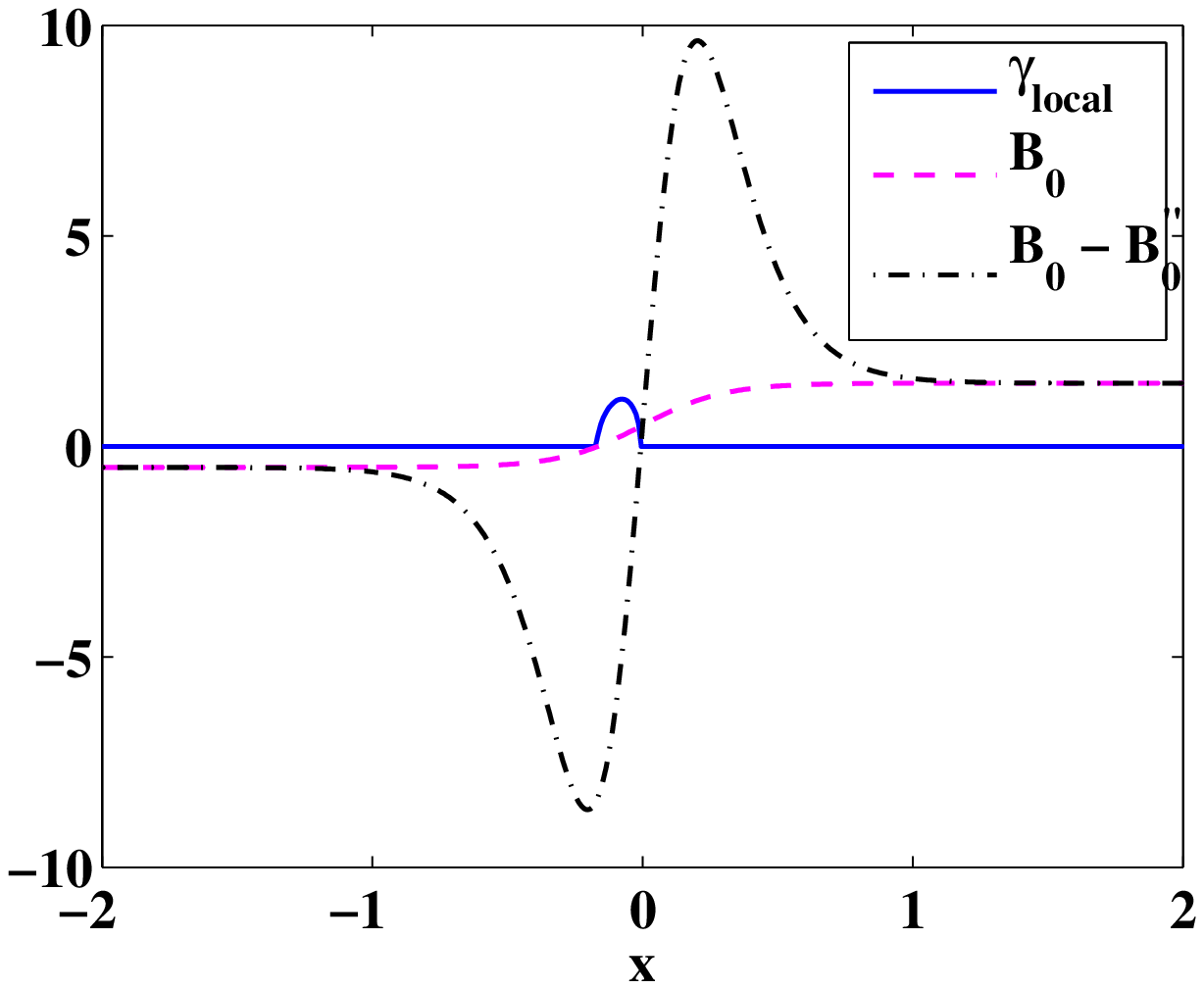}
\end{center}
\caption{}
\label{Fig4}
\end{figure}

\begin{figure}
\begin{center}
\includegraphics[width=18.cm,height=12.cm]{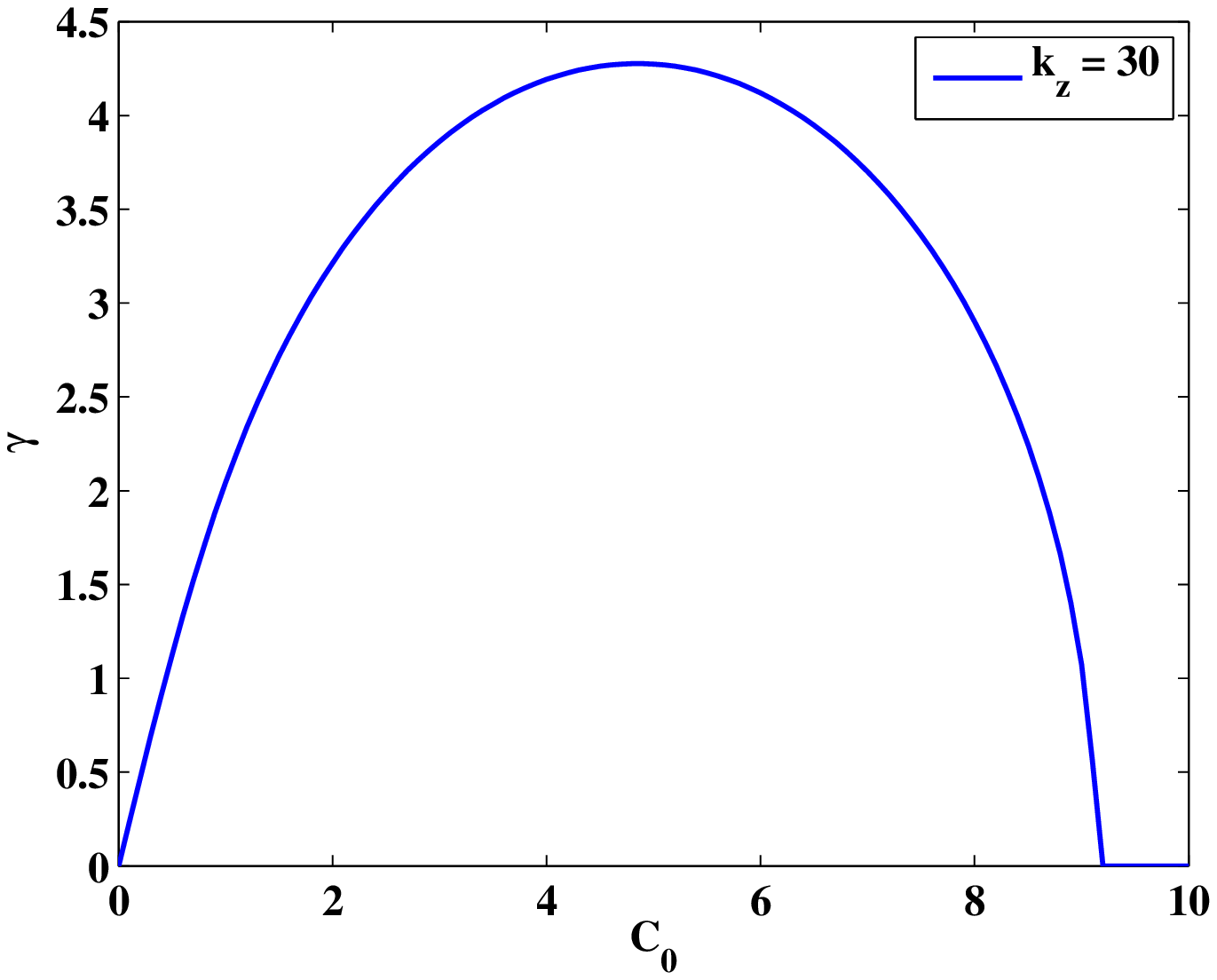}
\end{center}
\caption{}
\label{Fig5}
\end{figure}

In order to understand the role of $C_0$, we write $B_0 = B_0 +C_0$ in the above dispersion relation and obtain,
\begin{eqnarray}
\omega^2 (1+k_0^2) = k_z^2(B_0-B_0^{\prime\prime})(k_0^2 B_0+B_0^{\prime\prime}) + k_z^2 C_0 [2 k_0^2 B_0 + k_0^2 C_0 + B_0 (1-k_0^2)]
\label{loc_dp_c}
\end{eqnarray}
In the limit $k_z^2 >> 1$,
\begin{eqnarray}
\omega^2 (1+k_0^2) = k_z^2 k_0^2 (B_0+C_0)(B_0-B_0^{\prime\prime}+C_0)
\label{loc_dp_c_kz}
\end{eqnarray}
Below we discuss some cases for different values of $C_0$: 
 \begin{itemize}
  \item \underline{\it Case (i):} For $C_0 = 0$, the dispersion relation reduces to, 
\begin{eqnarray}
\omega^2 (1+k_0^2) = k_z^2 k_0^2 B_0(B_0-B_0^{\prime\prime}) = k_z^2 k_0^2 (B_0^2 - B_0 B_0^{\prime\prime})
\end{eqnarray} 
 For profiles like tanh, where the quantity $B_0 B_0^{''} < 0$ everywhere, there is no local instability. It is the case of purely 
 tearing instability as the condition $B_0 B_0^{''} > 0$ for non-tearing mode to be present is not satisfied.
 \item \underline{\it Case (ii):} For $0 < C_0  <1$, the quantity $B_0 B_0^{''}$ is positive in some region and negative in the other. 
 In this case local instability might be present. In Fig.~\ref{Fig4} we show that in the region where $B_0 B_0^{''} >0$, the local 
 instability is present. This is the case of mixed tearing and non-tearing modes.
 \item \underline{\it Case (iii):} For $C_0 > 1$, the null-point is removed, $B_0 B_0^{''} > 0$ everywhere, 
 no tearing instability. This is purely non-tearing instability.
 \item \underline{\it Case (iv):} For $C_0 > max(max|B_0|,max|B_0-B_0^{\prime\prime}|)$, the RHS becomes positive, and again there 
 is no instability. We show in Fig.~\ref{Fig5} that for certain value of $C_0$ there is no instability, this value $C_0 = C_0^{stable}$ 
 as pointed out earlier in this section. At this value the quantity $B_0-B_0^{\prime\prime}$ has definite sign (positive). This infers 
 that for instability the quantity $B_0-B_0^{\prime\prime}$ should change sign. Here, $C_0^{stable} = max (|B_0-B_0^{\prime\prime}|) \sim 9.1371$.
  \end{itemize}
 
\section{Nonlinear Simulations}

The coupled set of 2D EMHD evolution Eqns. (Eqn.(\ref{2d_emhd}) in Section II) can be expressed in the form of generalized continuity 
equations with source terms which have been evolved in slab geometry. A package of subroutines LCPFCT \cite{boris2} has been used which uses the 
flux corrected transport algorithm \cite{boris1}. The output of LCPFCT gives $\nabla^2 {f} - {f}$ where, $f \equiv b, \psi$ at each time step. A 
2D Helmholtz solver is employed for evaluating $b$ and $\psi$ at the updated time. The components of velocity and magnetic fields can be 
calculated using the relations, $\vec{B} = b\hat{y}+\hat{y}\times\nabla\psi$ and 
$\vec{v_e} = \hat{y}\times\nabla b-\hat{y}\nabla^2\psi$. 
An equilibrium configuration given as, $\psi_0(x) = \epsilon \log \{\cosh(x/\epsilon)\} - c_0 x$ and $b_0 = const ( = 0)$ has been 
chosen that would describe an equilibrium magnetic field $\vec{B_0} = \hat{z}(\tanh(x/\epsilon) + C_0)$. The equilibrium has been 
evolved against very low amplitude 
random numerical noise. We carry out the three simulation runs for $C_0 = 0.0, 0.5$ and $1.0$ keeping all the other parameters same. 
The evolution of total energy was tracked throughout the time of evolution in all the simulation runs to ascertain the 
accuracy.

\begin{figure}
\begin{center}
\includegraphics[width=18.cm,height=12.cm]{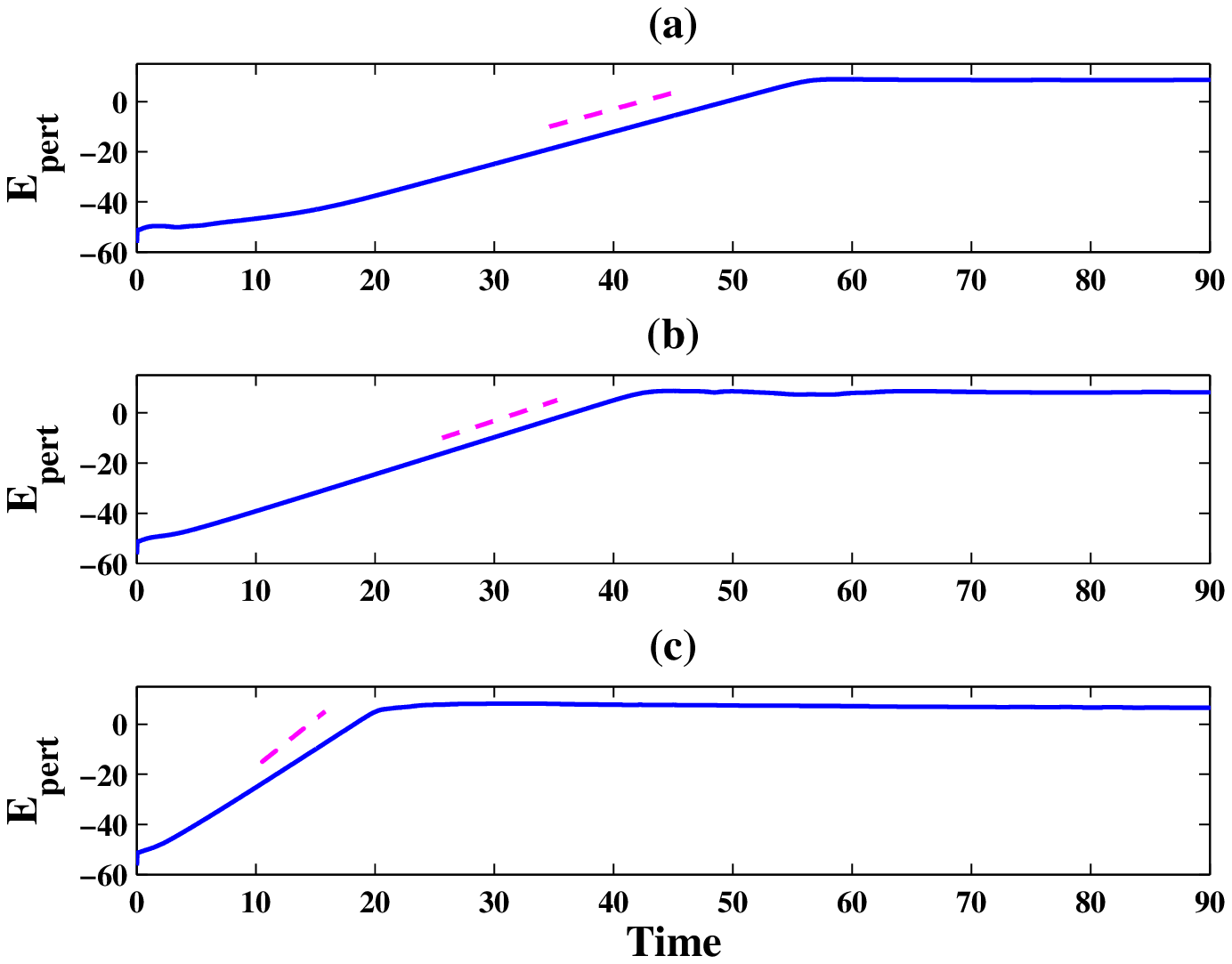}
\end{center}
\caption{}
\label{en_pert}
\end{figure}

In Fig. \ref{en_pert} we show the evolution of the perturbed energy of the system for $C_0$ = 0.0, 0.5, and 1.0. During the initial
phase of the simulation the total perturbed energy increases exponentially. In the semilog plot of Fig. \ref{en_pert} this can be seen
initially where the curve is a straight line. The slope of this line matches closely with twice the maximum growth rate 
obtained in Sec. III for each of the distinct values of $C_0$ . The dashed line shown alongside the simulation curve has
twice the slope corresponding to the analytical value of the maximum growth rate. As the amplitude of the perturbed field increases, 
the nonlinear effects become important in the simulation resulting in the saturation of the perturbed energy seen at the later stage.

\begin{figure}
\begin{center}
\includegraphics[width=20.cm,height=12.cm]{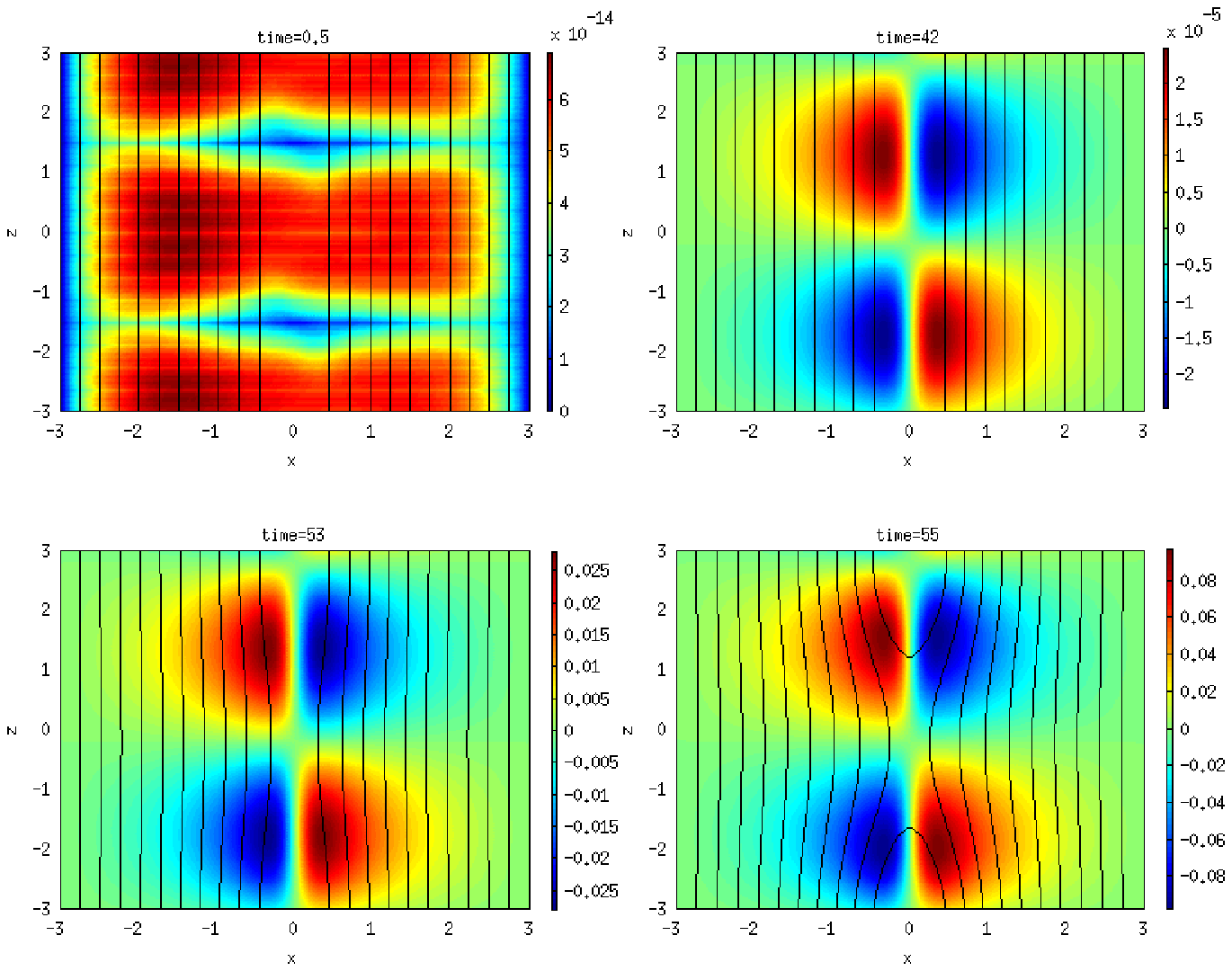}
\end{center}
\caption{}
\label{cont_0.0}
\end{figure}

In Fig.~\ref{cont_0.0} we show the contour plots of out of plane magnetic field $b$ for the case $C_0 = 0$. Also plotted are the 
contours of magnetic flux function $\psi$. It can be seen in the figure that the magnetic field lines which are straight initially 
get deformed at later times and show the reconnection of field lines. Consequently, an island type of structure is formed. 
The contour structure of $b$ is random initially which are initial low amplitude numerical noises. Later the field evolves into a 
quadrupole structure formed at X - point. This observed behavior is typical for tearing instability seen elsewhere also \cite{sarto_califano}. 
The non-tearing mode will not be present here, as pointed out in linear analysis carried out in the previous section.

\begin{figure}
\begin{center}
\includegraphics[width=20.cm,height=12.cm]{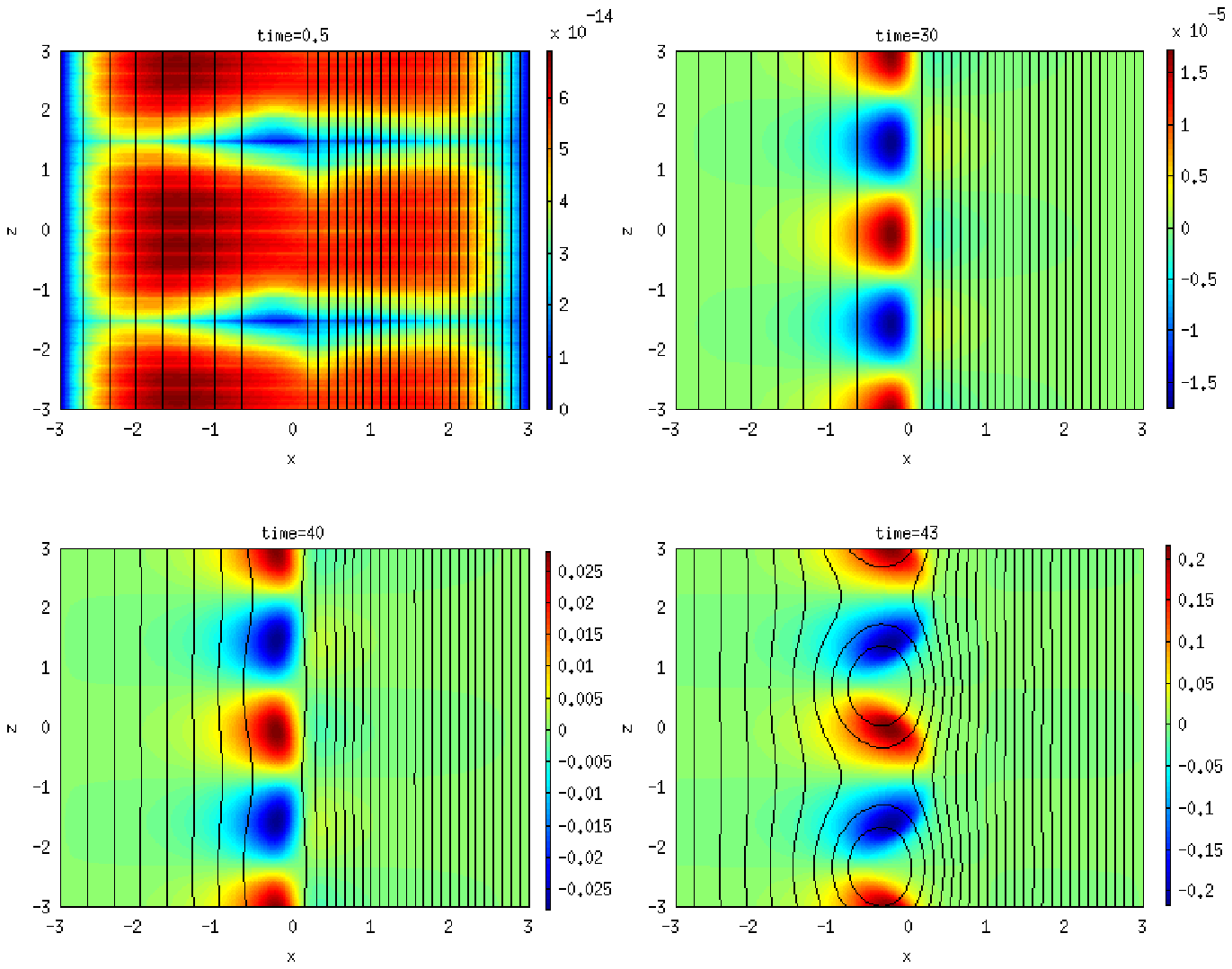}
\end{center}
\caption{}
\label{cont_0.5}
\end{figure}

We now discuss the next simulation simulation run with $C_0 = 0.5$. Here, the non-tearing mode will also be present. 
In Fig.~\ref{cont_0.5}, we show the evolution of $\psi$ and $b$. The initially straight field lines evolve and show the 
reconnection of field lines again. Here we see the formation of two islands because the box size permits two wavelengths of the 
fastest growing mode. But unlike the case for $C_0$, the islands formed here are asymmetric. The location of island is at $x \sim -~0.1648$, 
the location of null-point. The reason of asymmetry is due to the asymmetry in the strength of magnetic field on the two sides of the 
null-point. The asymmetry is observed in evolution of $b$ also, the quadrupole formed is asymmetric. These findings are in
accordance with asymmetric reconnections \cite{ass_reconn}. Asymmetric reconnections are expected to occur at magnetopause, 
where density and magnetic field strength on two sides of dissipation region are different \cite{magnetopause}.

\begin{figure}
\begin{center}
\includegraphics[width=20.cm,height=12.cm]{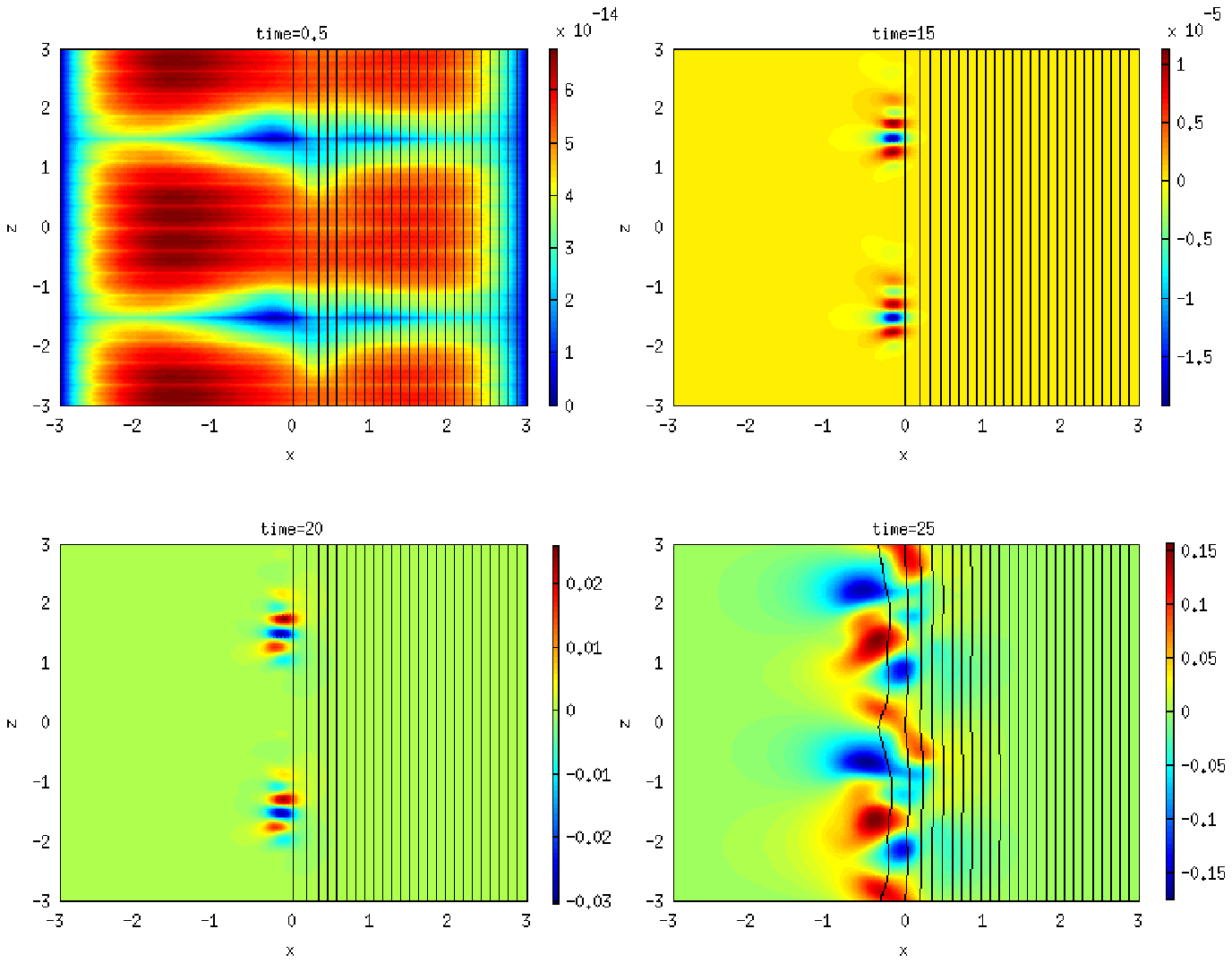}
\end{center}
\caption{}
\label{cont_1.0}
\end{figure}

Eventually, we study the case $C_0 = 1.0$. For this case, there is no null-point in the magnetic field profile, and hence 
tearing instability will not be present. This is confirmed in the evolution of $\psi$ in Fig.~\ref{cont_1.0}. Here, initially 
straight field lines show deformation at later times, but no reconnection is observed. In the final state, the magnetic filed line 
form a channel like structure. The evolution of $b$ variable is also completely different from that of tearing instability. Here, 
instead of a quadrupole, we see the localized small scale patterns which decay later.

\section{Summary} 
We have investigated tearing and surface preserving modes of Electron Magnetohydrodynamics (EMHD) in a current layer. Both 
linear as well as nonlinear studies have been carried out. The tearing instability breaks the magnetic field lines in the 
presence of electron inertia and leads to magnetic reconnection. Whereas, surface preserving mode unlike the tearing mode 
preserves the magnetic flux surface. We have called this mode non-tearing mode.

Linear perturbation analysis for a tangent hyperbolic profile of equilibrium magnetic field shows the existence of tearing mode. 
To study the non-tearing mode we add a uniform magnetic field $C_0$. Presence of $C_0$ satisfies the condition of non-tearing 
mode $B_0 B_0^{\prime\prime} > 0$. Tearing mode is a non-local mode and requires the null-line in the magnetic field. While, the 
non-tearing mode is a local mode and does not require null-line in the magnetic field. For tearing mode, the growth rate curve 
has non-local behaviour, all the local modes are stable. While, for non-tearing mode the growth rate curve shows the asymptotic 
behaviour in the local region. The change in sign for quantity $B_0-B_0^{\prime\prime}$, where $B_0$ is the equilibrium magnetic 
field, is necessary for any of these instabilities to exist.

For pure tearing case $C_0 = 0.0$, we observe the formation of magnetic island at the null-line due to the magnetic reconnection 
in the nonlinear state. The out-of-plane magnetic field shows the formation of quadrupole. These observations are typical for 
tearing instability. In the simulation with $C_0 = 0.5$, when both tearing and non-tearing mode are present we see the island 
formed is asymmetric. The quadrupole pattern in out-of-plane magnetic field is also asymmetric. These finding are in accordance 
with asymmetric reconnections in which magnetic field on two sides of dissipation region is asymmetric. In the case $C_0 = 1.0$, 
when there is no null-line present in the magnetic field, the tearing mode will not be present. In this case we do not observe 
the island structure, instead, we observe a channel-like pattern in the nonlinear state. 

\newpage
\centerline{FIGURE CAPTION}
\begin{itemize}

\item[Fig.1]
The schematic describes tearing and bending modes depending upon the orientation of perturbations relative to one dimensional 
equilibrium magnetic field $B_0(x)\hat{z}$. This magnetic field is created by an equilibrium electron flow $v_0(x)\hat{y}$ sheared along 
$x$ direction. Perturbations lying in the vertical plane, containing magnetic field with a null-line, give rise to tearing instability. 
When the angle of perturbations is changed to lie in the horizontal plane of shear and flow, the instability changes from tearing type to 
bending type. Both the instabilities are driven by velocity shear or equivalently, current shear in system where electron dynamical 
response is only of relevance.

\item[Fig.2] 
The figure shows the equilibrium magnetic field profile, $B_0 = tanh(x/\epsilon)+C_0$. The null-point is located at $x = 0 $ 
for $C_0 = 0$. The null-point shifts to the left for $C_0 = 0.5$ and vanishes for $C_0 = 1.0$. $2\epsilon$, being the shear width, 
remains same for all cases.  

\item[Fig.3] 
Plot of growth rate vs $k_z\epsilon$ for profiles given in Fig.1 with $\epsilon = 0.3$. Different curves are for 
different values of $C_0$. 

\item[Fig.4]
Surface plot of growth rate as a function of $k_z\epsilon$ and $C_0$.

\item[Fig.5]
Eigen function plots of pure tearing mode and of pure non-tearing mode in left and right panels, respectively. The other parameter 
values are $\epsilon = 0.3$ and $k_z = 1.0$

\item[Fig.6] 
This figure shows the growth rate curves obtained from non-local and local calculations for $\epsilon = 0.3$ and $C_0 = 1.0$. The growth rate 
from two curves are seen to match at large $k_z$ values. The reason of mismatch at small $k_z$ values is that the local analysis is not 
valid there. 

\item[Fig.7] 
The figure shows that the local modes are unstable in the region where $B_0B_0^{''} > 0$. Other parameters are $\epsilon = 0.3$, $k_z = 30$, 
$C_0 = 0.5$. $\gamma_{local}$ is the growth rate obtained from local analysis. Also plotted are $B_0$ and $B_0 - B_0^{''}$. 
$Max|B_0-B_0^{\prime\prime}| = 9.1371$.

\item[Fig.8] 
The local growth rate has been shown as a function of $C_0$. The other parameter values are $\epsilon = 0.3$ and $k_z = 30$.  

\item[Fig.9] 
The evolution of perturbed energy for $C_0$ = 0.0, 0.5, and 1.0 in subplots (a), (b), and (c), respectively. The dashed straight 
lines shown alongside each of the plots have been drawn with a slope of 2$\gamma_l$, where $\gamma_l$ is the linear growth rate of 
the system. The value of $\gamma_l$ is 0.65, 0.78, and 1.9 for $C_0$ = 0, 0.5, and 1.0, respectively.

\item[Fig.10] 
Shaded isocontours of the out of plane magnetic field at various times for the nonlinear simulation of $C_0 = 0.0$ case. 
Superimposed (solid lines) are the isocontours of magnetic flux function $\psi$.

\item[Fig.11]
Shaded isocontours of the out of plane magnetic field at various times for the nonlinear simulation of $C_0 = 0.5$ case. 
Superimposed (solid lines) are the isocontours of magnetic flux function $\psi$.

\item[Fig.12] 
Shaded isocontours of the out of plane magnetic field at various times for the nonlinear simulation of $C_0 = 1.0$ case. 
Superimposed (solid lines) are the isocontours of magnetic flux function $\psi$.

\end{itemize}

\newpage


\newpage
\begin{thebibliography}{99}

\bibitem{kingsep}
A. S. Kingsep, K. V. Chukbar and V. V. Yankov, in {\it Reviews of Plasma Physics} 
(Consultants Bureau, New York, 1990), 
Vol. 16 and references therein.

\bibitem{zp1}
A. A. Chernov and V. V. Yankov, Soviet Journal of Plasma Physics {\bf 8}, 522 (1982). 
\bibitem{zp2}
D. D. Ryutov, M. S. Derzon, and M. K. Matzen, Review of Modern Physics {\bf 72}, 000167 (2000). 

\bibitem{gordeev}
S. V. Basova, S. A. Varentsova, A. V. Gordeev, A. V. Gulin, and V. Yu. Shuvaev, 
Sov. J. Plasma Phys. {\bf 17}, 362 (1991). 
\bibitem{reconn1}
S. V. Bulanov, F. Pegoraro, and A.S. Sakharov, Phys. Fluids B {\bf 4}, 2499 (1992).
\bibitem{reconn2}
K. Avinash, S. Bulanov, T. Esirkepov, P. Kaw, F. Pegoraro, P. V. Sasorov, and A. Sen, Phys. Plasmas {\bf 5}, 2849 (1998).
\bibitem{reconn3}
F. Califano, N. Attico, F. Pegoraro, G. Bertin, and S. V. Bulanov, Phys. Rev. Lett. {\bf 86}, 5293 (2001).
\bibitem{reconn4}
L. I. Rudakov, and J. D. Huba, Phys. Rev. Lett. {\bf 89}, 095002 (2002). 
\bibitem{reconn5}
J. F. Drake {\it et al.}, Science {\bf 299}, 873 (2003).
\bibitem{reconn6}
L. Chacon, A. N. Simakov, and A. Zocco, Phys. Rev. Lett. {\bf 99}, 235001 (2007).
 \bibitem{reconn7}
 N. Jain and S. Sharma, Phys. Plasmas {\bf 16}, 050704 (2009).
 \bibitem{reconn8}
H. Che, J. F. Drake, and M. Swisdak, Nature{\bf 474}, 184 (2011). doi: 10.1038.

\bibitem{tabak}
M.~Tabak, J.~Hammer, M.~E. Glinsky, W.~L. Kruer, S.~C. Wilks, J.~Woodworth,
  E.~M. Campbell, M.~D.Perry and R.~J. Mason,
\newblock {\em Phys. Plasmas} {\bf 1}, 1626 (1994).
\bibitem{hain_mulsar}
S. Hain and P. Mulsar, Physical Review Letters {\bf 86}, 1015 (2001). 

\bibitem{pos1}
A. Fruchtman, A. A. Ivanov, and A. S. Kingsep, Phys. of Plasmas 
{\bf 5}, 1133 1998). 
\bibitem{pos2}
R. Shpitalnik {\it et al.}, Phys. Plasmas {\bf 5}, 792 (1998). 
\bibitem{pos3}
M. Sarfaty {\it et. al.}, Phys. Plasmas {\bf 2}, 2583 (1995). 
\bibitem{inter-planetary}
A. Teste and G. K. Perks, Phys. Rev. Lett. {\bf 102}, 075003 (2009).

\bibitem{tear_bend}
F. Califano, R. Prandi, F. Pegoraro, and S. V. Bulanov, Phys. Plasmas {\bf 16}, 2332 (1999). 
\bibitem{das_kaw_kh}
 A. Das and P. Kaw, Phys. Plasmas {\bf 8}, 4518 (2001).
\bibitem{gaur1}
 G. Gaur, S. Sundar, S. K. Yadav, A. Das, P. Kaw and S. Sharma, Phys. Plasmas {\bf 16}, 072310 (2009). 
\bibitem{pukhov}
S. V. Bulanov, M. Lontano, T. Zh. Esirkepov, F. Pegoraro, and A. M. Pukhov, Phys. Rev. Lett. {\bf 76}, 3562 (1996). 
\bibitem{drake_99}
D. Biskamp, E. Schwarz, A. Zeiler, A. Celani, and J. F. Drake, Phys. Plasmas {\bf 6}, 751 (1999).
\bibitem{jain_pla}
N. Jain, A. Das, P. Kaw, and S. Sengupta, Phys. Lett. A {\bf 363}, 125 (2007).
 \bibitem{jain_kink_lin} 
 N. Jain, A. Das and P. Kaw, Phys. Plasmas {\bf 11}, 4390 (2004).
\bibitem{lukin}
V. S. Lukin, Phys. Plasmas {\bf 16}, 122105 (2009).
 \bibitem{gaur2}
 G. Gaur and A. Das, Phys. Plasmas {\bf 19}, 072103, (2012). 
\bibitem{harris}
E. G. Harris, Nuovo Cimento {\bf 23}, 115 (1962).
 \bibitem{sarto_califano}
D. Del Sarto, F. Califano, and F. Pegoraro, Phys. Plasmas {\bf 12}, 012317 (2005).
 \bibitem{boris1}
 J. P . Boris and D. L. Book, Methods Compt. Phys. {\bf 16}, 76 (1976).
 \bibitem{boris2}
 J. P. Boris, {\it Flux Corrected Transport Modules for Generalized Continuity Equations,} (NRL Memorandom Report 3237, Naval Research 
Laboratory, Washington DC), 1976. 
\bibitem{ass_reconn}
P. A. Cassak and M. A. Shay, Phys. Plasmas {\bf 14}, 102114 (2007); P. A. Cassak and M. A. Shay, {\it ibid} {\bf 16}, 055704 (2009).
\bibitem{magnetopause}
T. D. Phan and G. Paschmann, J. Geophys. Res. {\bf 101}, 7801 (1996), doi: 10.1029/95JA03752; 
H. C. Ku and D. G. Sibeck, {\it ibid} {\bf 102}, 2243 (1997), doi: 10.1029/96JA03162.

\end{thebibliography}
 \end{document}